\documentclass[aps,twocolumn,showpacs,superscriptaddress]{revtex4-1}
\usepackage{amsmath,amssymb}
\usepackage{graphics,graphicx}
\usepackage{dcolumn,bm}
\usepackage{psfrag}
\usepackage{subfigure,makecell}

\newcommand{\cu}
{\affiliation{Department of Physics, University of Calcutta,
92 Acharya Prafulla Chandra Road, Kolkata 700009, India.}}

\begin{document}

\title{Zero temperature coarsening in Ising model with asymmetric second neighbour interaction  in  two dimensions}

\author{Pratik Mullick}%
\cu
\author{Parongama Sen}%
\cu

\normalsize  

\begin{abstract}

We consider the zero temperature coarsening in the Ising model in two dimensions where the spins interact within the Moore neighbourhood. 
The Hamiltonian is given by 
$H = - \sum_{<i,j>}{S_iS_j} - \kappa \sum_{<i,j'>}{S_iS_{j'}}$  where the two terms are for the first neighbours and second neighbours respectively and $\kappa \geq 0$. 
The freezing phenomena,
already noted in two dimensions for $\kappa=0$, is  seen to be present for any $\kappa$. However, the frozen states show more complicated structure 
as $\kappa$ is increased; e.g. local anti-ferromagnetic motifs can exist for $\kappa>2$.   
Finite sized systems also show the existence of an iso-energetic active phase  for 
$\kappa > 2$, which  vanishes in the thermodynamic limit. 
The persistence probability shows universal behaviour for $\kappa>0$, however it is clearly different from the $\kappa=0$ results
when non-homogeneous initial condition is considered.
Exit probability 
shows universal behaviour for all $\kappa \geq 0$. 
The results are compared with other models in two dimensions having interactions beyond the first neighbour.

\end{abstract}


\maketitle

\section{Introduction}

Non-equilibrium dynamics in various interacting particle systems has emerged as one of the most studied  topics 
over the past few decades. 
In physics, 
many non-equilibrium phenomena can be described in terms of interacting spin systems where the evolution of the 
system may either be dictated by some dynamical rules or in terms of a Hamiltonian. 
In classical spin systems like the Ising model, which have no intrinsic dynamics, thermal noise or other driving forces can introduce 
dynamics in the system. 
Interestingly, even without the presence of thermal noise, one can study the evolution of the system starting from 
configurations far from equilibrium which is actually a non-equilibrium dynamical phenomena. 
In fact,  the behaviour of subcritical
dynamics is universal and essentially the same as that in zero-temperature dynamics. For this reason a large number of studies have been
concentrated on zero-temperature dynamics \cite{bray,krapbook}.


At zero temperature, in the Ising model, 
the coarsening phenomena is surface tension-driven \cite{bray,krapbook} with energy costs taking place at the domain interfaces \cite{corberi}. 
One of the main features in coarsening at low temperature is the domain growth phenomena. Following dynamical scaling
hypothesis, the characteristic length scale $L(t)$ grows with time as $t^{\frac{1}{z}}$ \cite{bray,halperin}. 
In the infinite system, as domains of up and down spins both grow at this rate the system may evolve for ever. However, for finite systems, a random fluctuation may drive the system to the uniform state of all up or all down spins which are the equilibrium ground states. 
Naively, one would expect  this to happen all the time. However, it has been shown that in dimensions greater than one, the
spins can be locked in frozen phases such that although the state has higher energy, the spins at the interfaces cannot flip as that
increases their energy \cite{lipowski,spirin63,spirin65,barros,prl-redner}. This is when one considers single spin flip energy minimising dynamics, e.g., Glauber dynamics, where the evolution does not conserve the order parameter. In two dimensions, such frozen 
configurations occur as the domains form a striped pattern. Staircase like patterns can emerge when interactions with a longer range is considered. Comparison with critical percolation helps in obtaining some exact estimates of the freezing probability \cite{barros, prl-redner}.  In three dimensions, one has iso-energy ``blinkers'' which means the spin flips take 
place but the energy remains same \cite{spirin63,spirin65}. Only by introducing noise or external fields can the system escape from such frozen states.  

Another interesting phenomena in coarsening processes is persistence; the probability that a field has not changed sign till
a certain time
has been shown to have a power law decay in time as $P(t)\sim t^{-\theta}$ 
in many systems \cite{derrida1,satya3,stauffer1}. 
The corresponding exponent $\theta$ is not related to any known static or dynamic exponent. 
In spin models, persistence probability is estimated as the probability that a spin has not flipped till time $t$.    
In the Ising model, exact result for the one dimensional case exists while  in higher dimensions, only approximate results
are available \cite{derrida1,satya3,stauffer1,derrida2,bennaim,jain,blanchard,subir1,subir2,pratik}. 

Recently, coarsening of many systems with deviation from the completely disordered state has also been considered. 
In spin models this implies that one can start with $x$ fraction of up spins, with $x$ not necessarily equal to 0.5.
While 
the freezing phenomena has been shown to disappear for $x\ne 0.5$ in the thermodynamic limit, the quantity of interest is the exit probability   $E(x)$ 
which is  the probability to reach the configuration with all spins in the up state. 
In one dimension, in the Ising model with only nearest neighbour interaction, $E(x) = x$. In higher dimensions, $E(x)$ is not only non-linear, there is also strong system size dependence which suggests a step function like behaviour in the thermodynamic limit \cite{pratik,parna-arxiv}.
While calculating $E(x)$ at $x = 0.5$ in two dimensions, care is taken to discard the configurations which do not reach the equilibrium ground state.
For $x\ne 0.5$, three types of persistence probabilities can be defined, $P_{min}$ (persistence probability for the type of spin with initial fraction $x<0.5$), $P_{maj}$ (persistence probability for the type of spin with initial fraction $x>0.5$) and $P_{total}$ \cite{bennaim}. 
Numerical results show that $P_{min}$ vanishes as $t^{-\gamma}\exp[-(t/\tau)]$, while the persistence probabilities for the
majority and total spins decay to high saturation values in two dimensions \cite{pratik}. The exponents show systematic dependence on $x$.

When the range of the interaction is extended, several striking differences are noted in the dynamical behaviour
\cite{prl-redner,redkrap,das-barma,ps-sdg,chandra,chandra-sdg}. Freezing patterns may be affected for models with both competing and non-competing anisotropic second nearest neighbour interactions. The nature of persistence probability and domain growth dynamics change considerably compared  to the results of the nearest
neighbour Ising model in two dimensions when a competing anti-ferromagnetic interaction is considered \cite{chandra-sdg,castellano,parnajstat}. Exit probability (EP) for Ising model shows non-linear variations when the range of interaction is increased even
in one dimension \cite{parnasoham,castellano,parnajstat}, although there is no system size dependence.



\begin{figure}
\begin{center}
\includegraphics[width=9cm]{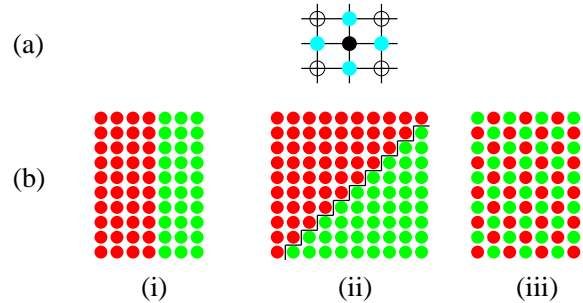}
 \caption{(a) Moore neighbourhood on a square lattice: the central point (black circle) has eight neighbouring points. The filled (empty) circles represent nearest (second nearest) neighbours. (b) Possible types of freezing states depending on the value of $\kappa$ are shown schematically: (i) Stripe state for any value of $\kappa$ (ii) Staircase shaped diagonal state for $\kappa>0$ (iii) Anti-ferromagnetic state for $\kappa>1$.  The red (green) dots represent up (down) spins.}
\label{fstates}
\end{center}
\end{figure}

In the present paper, we have considered the two dimensional
Ising model with second nearest neighbour interaction, as in a Moore neighbourhood (shown in Fig. \ref{fstates}(a)), where the interaction strengths of
the first and second nearest neighbours are different in general.
Our interest is to find out the freezing behaviour as a function of $\kappa$ as well as to estimate the 
persistence probability and exit probability for $\kappa\geq 0$. 

Although our interest in studying models with longer range interactions is primarily theoretical, there are materials which indeed have complex interactions beyond nearest neighbours. For example in SnTe, where such interactions exist, it is even possible that the further neighbour interaction dominates over the nearest neighbour one \cite{cwli}.

\section{Model and quantities calculated}

The Hamiltonian of the Ising model in two dimensions with next nearest neighbour interaction is given by
\begin{equation}
H = - J_1\sum_{<i,j>}{S_iS_j} - J_2\sum_{<i,j'>}{S_iS_{j'}},
\end{equation}
where $J_1$ and $J_2$ are the strengths of interaction for nearest neighbour and second nearest neighbour
respectively. We have taken $J_1=1$ and $\kappa = {J_2}$
is the parameter denoting the ratio of the interactions. Clearly, $\kappa = 0$ corresponds to the case of the  well studied nearest neighbour model.
We perform zero temperature quenching using Monte Carlo simulation. At every time step we pick up a spin at
random and flip it according to the configuration of its nearest and second nearest neighbours. Table \ref{table1} 
summarizes the possible states of the central spin (i.e. the randomly picked spin) with respect to the configuration of its nearest
and second nearest neighbours. Here ($N_N,N_{SN}$) indicates that there are $N_N$ nearest neighbours and $N_{SN}$
second nearest neighbours which are in the up state. It may be noted from the table, that for configurations (4,0) and (4,1) there is 
a possibility that the central spin becomes down for $\kappa>1$ and $\kappa>2$ respectively, favouring a local anti-ferromagnetic behaviour.

We calculate variation of the freezing probability $f_p$ for different system sizes and different values of
$\kappa$, keeping $x = 0.5$. Any configuration which does not reach the equilibrium ground state is 
termed a frozen state.
We have estimated the fraction of total spin flips $S_t$ as a function of time.
Since locally anti-ferromagnetic motifs can occur for $\kappa>1$, it is useful to calculate the density $A_t$
of such motifs (details to be discussed later).
Persistence probability and its variants, i.e., persistence for the total system $P_{total}$, for minority spins $P_{min}$
and for majority spins $P_{maj}$ were calculated numerically for different values of $\kappa$.
We also calculate the exit probability (EP) $E(x)$ as a function of $x$ for various values of $\kappa$ and different
system sizes.

\begingroup
\begin{table}

\label{table}
\begin{tabular}{|c|cc|c|cc|}
\Xhline{3\arrayrulewidth}
{}&{State of}&{}&{}&{State of}&{}\\
{($N_{N}$,$N_{SN}$)}&{the}&{}&{($N_{N}$,$N_{SN}$)}&{the}&{}\\
{}&{central}&{}&{}&{central}&{}\\
{}&{spin}&{}&{}&{spin}&{}\\
\Xhline{3\arrayrulewidth}
{}&{}&{}&{}&{}&{}\\
{(4,4)}&{$\uparrow$}&{}&{(3,4)}&{$\uparrow$}&{}\\
{}&{}&{}&{}&{}&{}\\
{(4,3)}&{$\uparrow$}&{}&{(3,3)}&{$\uparrow$}&{}\\
{}&{}&{}&{}&{}&{}\\
{(4,2)}&{$\uparrow$}&{}&{(3,2)}&{$\uparrow$}&{}\\
{}&{}&{}&{}&{}&{}\\
{}&{$\uparrow$}&{$(\kappa<2)$}&{}&{$\uparrow$}&{$(\kappa<1)$}\\
{(4,1)}&{$\circ$}&{$(\kappa = 2)$}&{(3,1)}&{$\circ$}&{$(\kappa = 1)$}\\
{}&{$\downarrow$}&{$(\kappa>2)$}&{}&{$\downarrow$}&{$(\kappa>1)$}\\
{}&{}&{}&{}&{}&{}\\
{}&{$\uparrow$}&{$(\kappa<1)$}&{}&{$\uparrow$}&{$(\kappa<0.5)$}\\
{(4,0)}&{$\circ$}&{$(\kappa = 1)$}&{(3,0)}&{$\circ$}&{$(\kappa = 0.5)$}\\
{}&{$\downarrow$}&{$(\kappa>1)$}&{}&{$\downarrow$}&{$(\kappa>0.5)$}\\
\Xhline{3\arrayrulewidth}
{}&{}&{}&{$N_N$}&{}&{}\\
{(2,4)}&{$\uparrow$}&{}&{For $\kappa = 0$}&{}&{}\\
\Xcline{4-6}{3\arrayrulewidth}
{}&{}&{}&{}&{}&{}\\
{(2,3)}&{$\uparrow$}&{}&{4}&{$\uparrow$}&{}\\
{}&{}&{}&{}&{}&{}\\
{(2,2)}&{$\circ$}&{}&{3}&{$\uparrow$}&{}\\
{}&{}&{}&{}&{}&{}\\
{(2,1)}&{$\downarrow$}&{}&{2}&{$\circ$}&{}\\
{}&{}&{}&{}&{}&{}\\
{(2,0)}&{$\downarrow$}&{}&{1}&{$\downarrow$}&{}\\
{}&{}&{}&{}&{}&{}\\
{}&{}&{}&{0}&{$\downarrow$}&{}\\
{}&{}&{}&{}&{}&{}\\
\Xhline{3\arrayrulewidth}
\end{tabular}
\caption{Possible states of the central spin with respect to $\kappa$ and the number of up spins in the nearest and second
nearest neighbour positions. $N_{N}$ and $N_{SN}$ indicate the number of up spins in the nearest and second nearest
neighbour positions respectively. The circles represent the undecided states when the probability of occurring of an up state
or down state is equal. The configurations for $N_{N} = 1$ and $0$ can be obtained from that of $N_{N} = 3$ and $4$ respectively, using the fact that the system is symmetric with respect to the position of up-down spins.}
\label{table1}
\end{table}
\endgroup

\begin{figure}
\includegraphics[width=8cm]{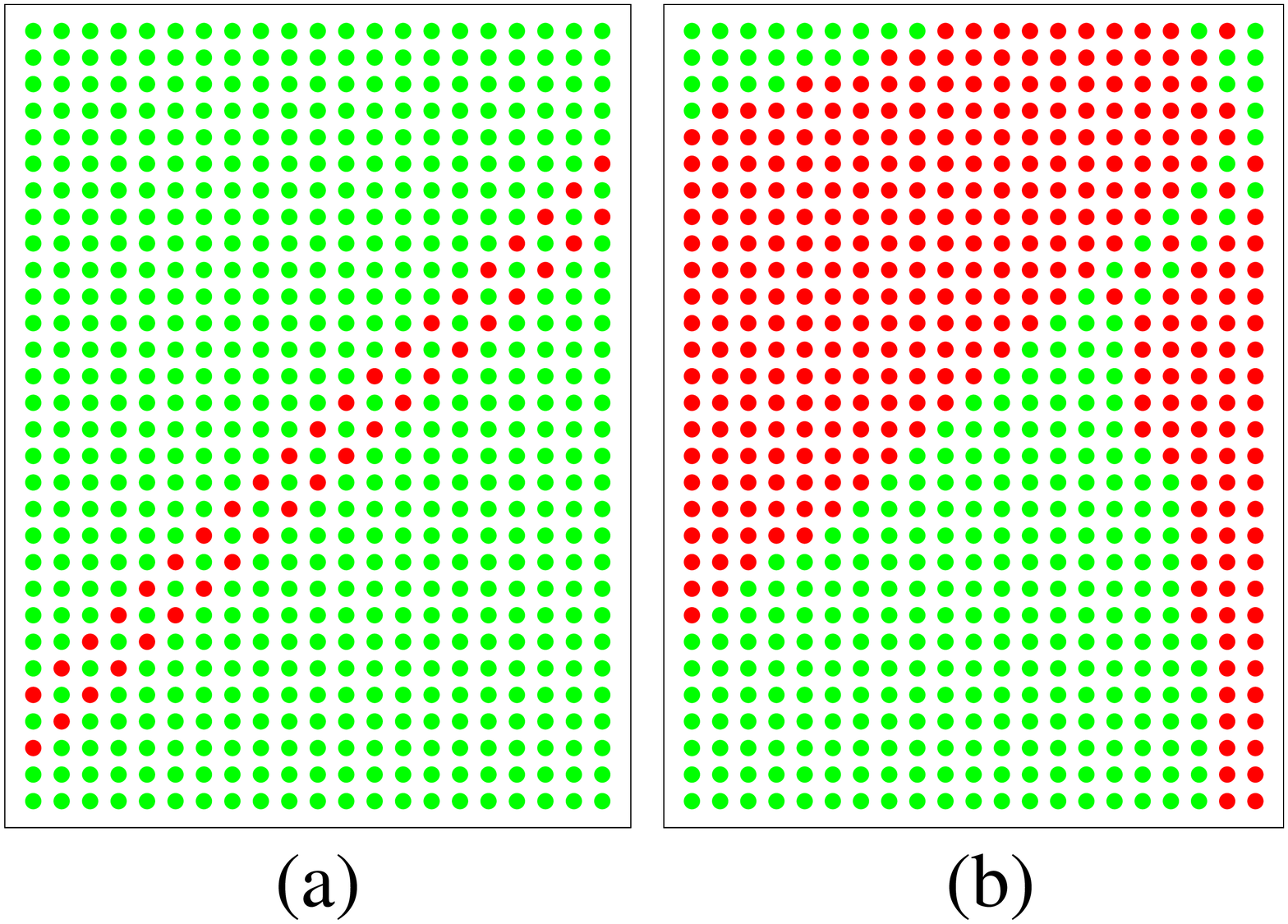}
 \caption{Snapshots of a section of $40 \times 40$ system at time $t = 3000$ with $\kappa = 2.25$ for two different configurations are shown. (a) shows a configuration with very short ranged anti-ferromagnetic order; this is an absorbing state. (b) shows a configuration with a combination of horizontal, vertical and diagonal stripe states, as well as some unstable spins. This particular configuration is identified as an iso-energetic active state as it does not reach a steady state within the observation time.}
\label{fstates810}
\end{figure}


\begin{figure}
\includegraphics[width=8cm]{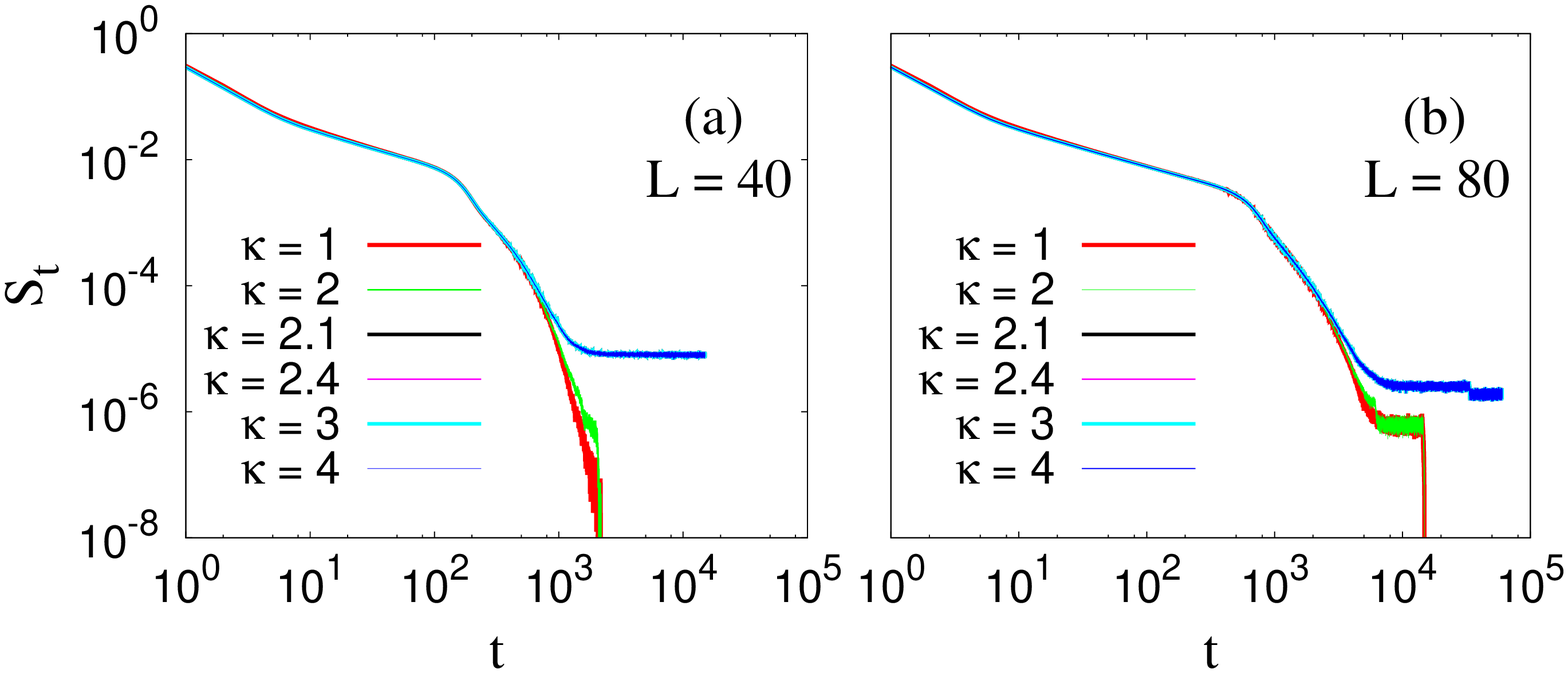}
 \caption{Variation of fraction of spin flips $S_t$ with time for six values of $\kappa$ and two different system sizes (a) $L=40$ and (b) $L = 80$ are shown. For the system size 40 $\times$ 40 the study was done up to time $t = 1.5 \times 10^4$, averaging over 5 $\times$ $10^4$ configurations and for the $80 \times 80$ system it was done up to time $t=6 \times 10^4$, taking average over $2 \times 10^4$ configurations. Only for $\kappa=$ 1 and 2, $S_t$ vanishes at large times.}
\label{spinflip}
\end{figure}

\begin{figure}
\includegraphics[width=8cm]{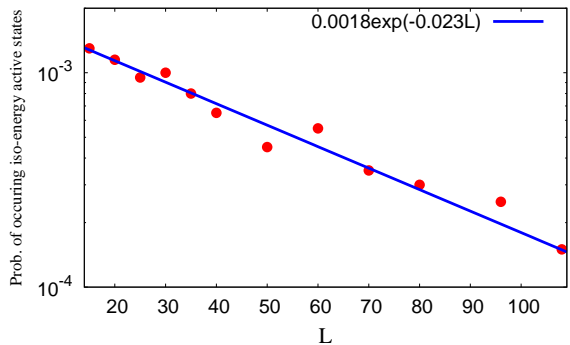}
 \caption{Probability of occurring iso-energetic active states as a function of system size $L$. The probability scales with $L$ as $a\exp(-bL)$.}
\label{active}
\end{figure}

\begin{figure}
\includegraphics[width=8cm]{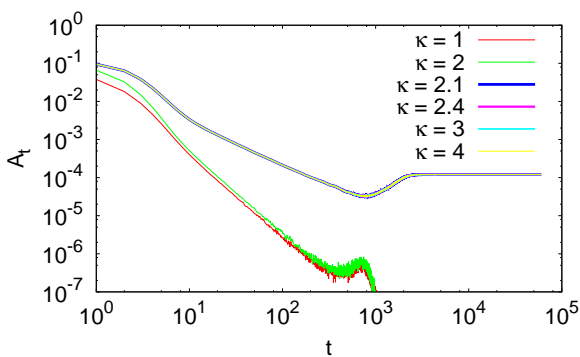}
 \caption{Fraction of anti-ferromagnetic motifs $A_t$ as a function of time for six different $\kappa$ values are shown. This study was done for a 80 $\times$ 80 system, averaging over $2 \times 10^4$ configurations. The data for $\kappa = 3$ and $4$ coincide. $A_t$ vanishes at
larger times for $\kappa=1,2$ only.}
\label{antiferro}
\end{figure}

\begin{figure}
\includegraphics[width=8cm]{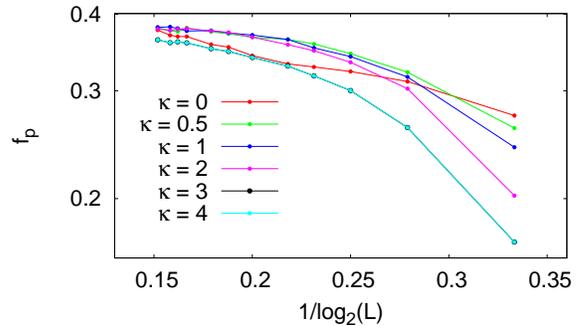}
 \caption{Variation of freezing probability with system size $L$ for six different values of $\kappa$ for $x = 0.5$ are shown. The data for $\kappa=3$ and 4 coincide.}
\label{fp1}
\end{figure}
\begin{figure}
\includegraphics[width=8cm]{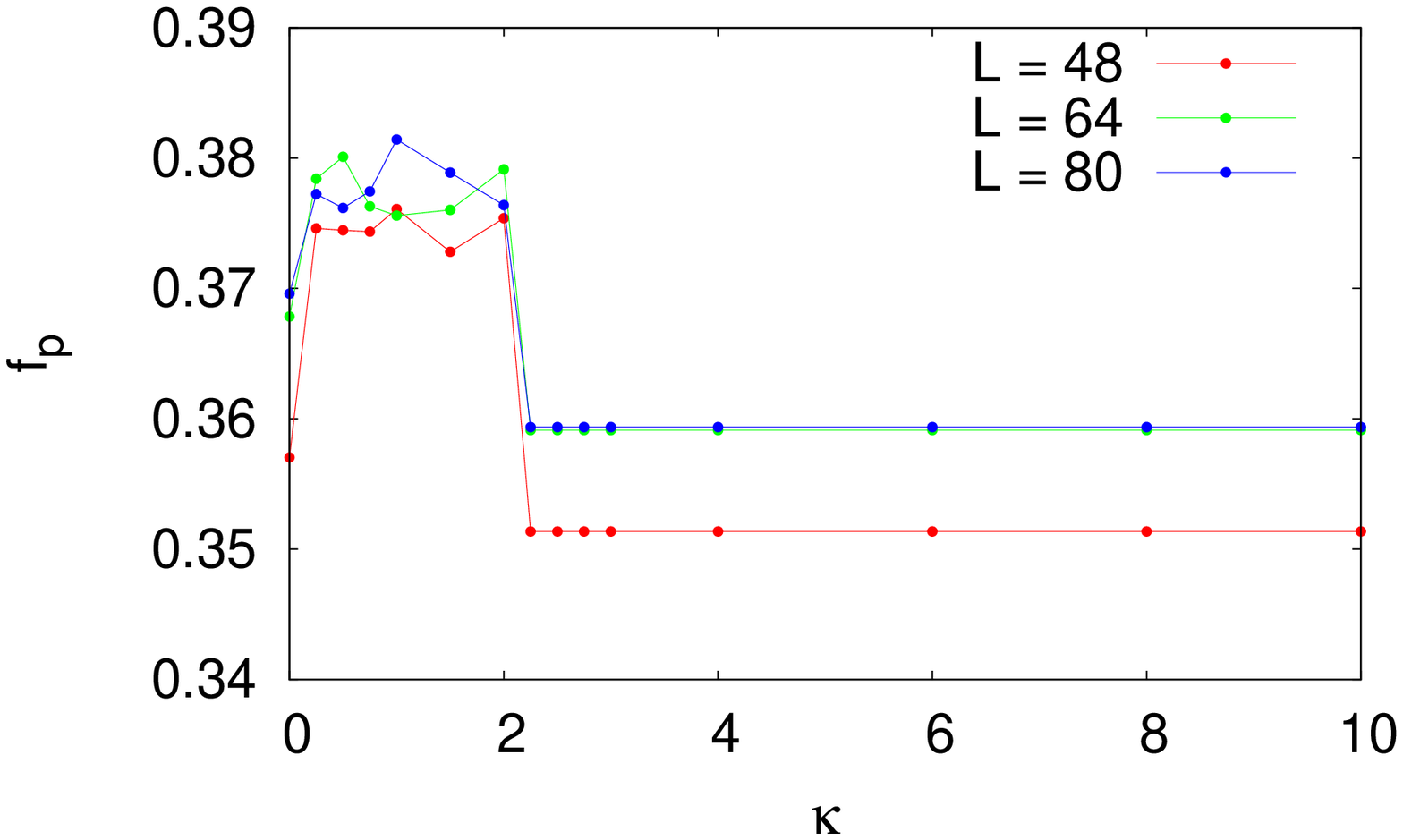}
 \caption{Variation of freezing probability with $\kappa$ for three different system sizes for $x = 0.5$ are shown.}
\label{fp2}
\end{figure}

The studies were performed on $L \times L$ square lattices. One Monte Carlo time step consisted of $L^2$ updates
and random asynchronous updating rule was used. Generally periodic boundary condition was imposed for the simulation
and helical boundary condition \cite{newman} was used only for the study of exit probability. Simulations were performed over $10^6$ configurations for $L \leq 16$
and $10^5$ configurations for $L \geq 20$.
EP was studied with $L \leq 64$ and simulations were performed for a maximum of $7000$
configurations. The maximum size simulated is $L = 108$.

\section{Results and analysis}

From Table \ref{table1} we note that significant changes in the dynamics takes place at 
$\kappa = 0.5, 1$ and  $2$. We have therefore kept $\kappa$ finite, $\leq 10$ in the simulations, which is sufficient to capture the effect
of $\kappa$ entirely.

\subsection{Freezing phenomena}

We first discuss the freezing phenomena in detail in this subsection. By freezing we specifically mean that the system reaches an absorbing state at a higher energy value compared to the ground state (all up/down spins). In Fig. \ref{fstates}b we have schematically shown
three possible types of freezing. From Table \ref{table1} it can be seen that staircase like diagonal states are stable
for $\kappa>0$, striped states are stable for any value of $\kappa$ and anti-ferromagnetic states are stable for
$\kappa>1$. The occurrence of fully anti-ferromagnetic state is extremely rare but it is possible that local anti-ferromagnetic patterns occur in the steady state. This may give rise to adjacent staircase like
interfaces, which has not been observed in earlier studies \cite{prl-redner}. An example is shown in Fig. \ref{fstates810}(a), obtained from the simulation of a $40 \times 40$ system with $\kappa = 2.25$.

To characterise the configuration of neigbouring spins of a particular spin one needs the total number of up spins in the nearest neighbour and second nearest neighbour positions denoted by $N_N$ and $N_{SN}$. As already mentioned in section II, a particular configuration is written as $(N_N,N_{SN})$. The configuration (2,3) and (2,1) gives rise to diagonal freezing states which are stable for any positive value of $\kappa$. The configuration (3,2) and (1,2) give rise to striped states which are stable for any value of $\kappa$. The anti-ferromagnetic states can occur due to the configurations (4,0) and (0,4) for $\kappa>1$. For $\kappa>2$, anti-ferromagnetic motifs may also occur for the configurations (4,1) and (0,3) (see Table \ref{table1}); these configurations have a higher occurrence probability compared to (4,0) and (0,4) configurations. Thus one can expect larger number of anti-ferromagnetic patterns for $\kappa>2$ and less for $1<\kappa<2$.
However, stable anti-ferromagnetic patterns were obtained only in the $\kappa>2$ region (checked for system sizes $L\leq 80$).

For $\kappa < 2$ we find that the system either reaches the equilibrium ground state or the frozen state
where the frozen  configurations  are either striped or resemble a  staircase pattern (Fig. 1b(i) and (ii)). 
However, on increasing $\kappa$, we find that certain configurations do not reach an absorbing state, one such example is shown in Fig. \ref{fstates810}(b). 
This is confirmed from the study of the fraction of spin flips $S_t$ as a function of time plotted in Fig. \ref{spinflip}; $S_t$ remains non-zero, however small, even at large times. For $\kappa<2$, $S_t$ decays with time and vanishes eventually, which indicates the absence of iso-energetic active states. For smaller system sizes the decay is smooth (Fig. \ref{spinflip}). However for larger systems (shown for $L = 80$) this decay eventually slows down and remains a constant over a period of time before abruptly going to zero. This indicates that the system gets locked in a metastable state and apparently a random fluctuation drives it to an absorbing state abruptly.
For $\kappa>2$, $S_t$ remains non-zero up to a very large time indicating the existence of iso-energetic active states. However, we observe that for larger system size, $S_t$ decays in steps for $\kappa>2$ (Fig. \ref{spinflip}), as it stays at a constant value for a long time and decreases to another constant value later. Hence it is difficult to predict whether asymptotically the system is driven to an absorbing state or not based on numerical simulations alone as one cannot predict what will happen at time $t \to \infty$.



For $\kappa>2$, even though the system remains active for some configurations where spins go on flipping, the energy remains constant, indicating these are iso-energetic active states.

To investigate whether the iso-energetic active states survive in the thermodynamic limit, we calculated the fraction of active states as a function of system size $L$. To do this we calculate the fraction of configurations for which the number of spin flips is non-zero even up to a very large time. The results are shown in Fig. \ref{active} which indicate that the fraction decays as $a\exp(-bL)$, with $a = 0.001806\pm9.94\times10^{-5}$ and $b = 0.02308\pm0.00169$. Clearly, the iso-energetic states vanish for $L \to \infty$. 
%

Previously we have discussed that the anti-ferromagnetic domains are likely to occur in the region $\kappa > 2$ with greater
probability than in the region $1 < \kappa < 2$.
In order to investigate quantitatively, we estimate the fraction of anti-ferromagnetic motifs. This was done by calculating the number of spins 
surrounded by four nearest neighbourDOI: 10.1103/PhysRevE.95.062703 spins of opposite orientation and dividing it by the total number of spins. From Fig. \ref{antiferro} one can see that for $\kappa>2$, the anti-ferromagnetic motifs can survive, while they disappear otherwise. It is interesting to note that there is a non-monotonic behaviour in both the regions $\kappa<2$ and $\kappa>2$.

\begin{figure}
\includegraphics[width=8cm]{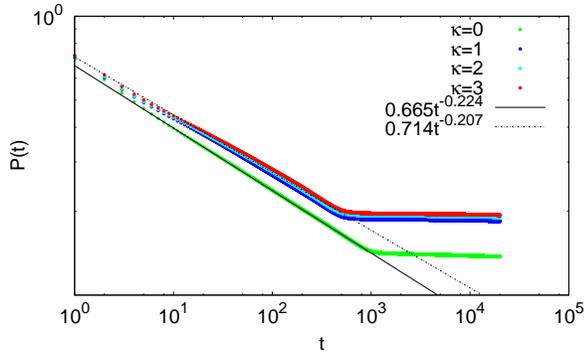}
 \caption{Variation of persistence probability $P_{total}$ for system size $64 \times 64$ for $x=0.5$ with $\kappa=$ 0, 1, 2 and 3 are shown. The data were fitted to the form $t^{-\theta}$.}
\label{per1}
\end{figure}

\begin{figure}
\includegraphics[width=8cm]{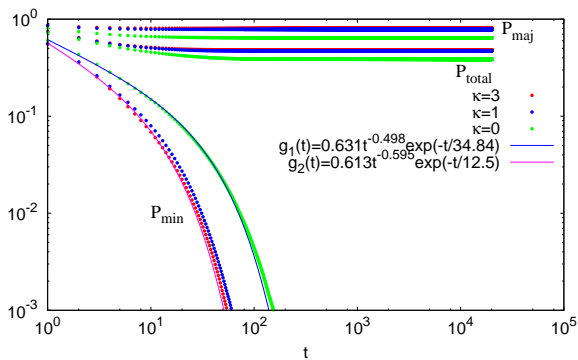}
 \caption{Three types of persistence probabilities as a function of time for a system size of $64 \times 64$ for $x=0.4$ with $\kappa=$ 0, 1 and 3 are shown. The data for $P_{min}$ fit to the functional form $t^{-\gamma}\exp{(-t/\tau)}$ with different values of $\gamma$ and $\tau$.}
\label{per}
\end{figure}

For a particular value of $\kappa$, freezing probability $f_p$ was calculated for different system sizes (Fig. \ref{fp1}).
It may be noted that $f_p$ includes all the states which do not reach ground state, i.e., it includes iso-energetic dynamic states as well. $f_p$ is seen to increase with the system size. Earlier studies have revealed
that for much larger system sizes, $f_p$ shows a tendency to decrease and the value for $L \rightarrow \infty$ is extrapolated from this region \cite{spirin65,spirin63}. We have not attempted such an extrapolation as the non-monotonic behaviour is not revealed for $L \leq 96$
considered in the present work and an extrapolation obviously leads to a higher value compared to 0.339 \cite{barros} for $\kappa=0$.
However, we can see an interesting dependence of $f_p$ on $\kappa$; apparently $f_p$ has two different constant values for $0<\kappa<2$ and $\kappa>2$. This is more clearly shown in Fig. \ref{fp2},
where the value of $f_p$ is $\approx 0.38$ for $0<\kappa\leq 2$ and drops abruptly to $\approx 0.36$ for $\kappa>2$ for the largest system size considered.


\subsection{Persistence probability}


The results obtained for persistence probability $P(t)$ with $\kappa>0$ is qualitatively similar to that for $\kappa=0$.
For the homogeneous case, i.e. $x=0.5$, the persistence probability $P(t)$ shows a power law decay with time as $t^{-\theta_0}$ for $\kappa=0$ and $t^{-\theta_{\kappa}}$ for $\kappa>0$ (Fig. \ref{per1}).
The values obtained from the simulations are $\theta_0 \approx 0.224$ and $\theta_{\kappa} \approx 0.207$ (independent of $\kappa>0$), having an error bar of the order of $10^{-5}$ and $10^{-4}$ respectively.  Both these values are sufficiently
close to the estimate of persistent exponent for two dimensional Ising model \cite{jain,blanchard,subir1,subir2,pratik}.

Results for the non-homogeneous case are shown in Fig. \ref{per}, where we show the variation for $P_{min}$, $P_{maj}$ and $P_{total}$ with time for $x=0.4$ for $\kappa=3.0$ and $\kappa=0$. 
$P_{min}$ is known to vanish as $t^{-\gamma}\exp[-(t/\tau)]$ for $\kappa=0$, with $\gamma=0.498 \pm 0.001$ and $\tau=34.84 \pm 0.21$.
While the same form is valid for $\kappa>0$, the values of $\gamma$ and $\tau$ differ considerably, e.g. for $\kappa=3$, $\gamma=0.595 \pm 0.008$
and $\tau=12.5 \pm 0.344$. However, $\gamma$ and $\tau$ have universal values for $\kappa>0$.

We find that the saturation values attained by the three persistence probabilities $P(t)$ (for the homogeneous case), $P_{maj}$ and $P_{total}$ (for the non-homogeneous case) are considerably larger for $\kappa>0$ compared to those for $\kappa = 0$. This is explained by the fact that a larger number of neighbours provides a greater stability to the spins, resulting in a larger saturation value of these persistence probabilities for $\kappa>0$. (Fig. \ref{per1},\ref{per}).


\subsection{Exit probability}

\begingroup
\begin{table}

\begin{tabular}{|c|c|c|c|}
\Xhline{3\arrayrulewidth}
{$\kappa$}&{$\nu$}&{$\lambda$}&{$\epsilon$}\\
\Xhline{3\arrayrulewidth}
{0}&{1.259}&{1.06}&{4.44$\times10^{-4}$}\\
{0.1}&{1.285}&{1.286}&{5.37$\times10^{-4}$}\\
{0.25}&{1.309}&{1.343}&{4.41$\times10^{-4}$}\\
{0.5}&{1.29}&{1.287}&{4.59$\times10^{-4}$}\\
{0.75}&{1.204}&{1.001}&{1.44$\times10^{-3}$}\\
{0.8}&{1.267}&{1.246}&{5.6$\times10^{-3}$}\\
{0.9}&{1.265}&{1.228}&{5.59$\times10^{-3}$}\\
{1}&{1.299}&{1.334}&{4.7$\times10^{-4}$}\\
{1.1}&{1.291}&{1.337}&{5.07$\times10^{-4}$}\\
{1.5}&{1.27}&{1.296}&{4.58$\times10^{-4}$}\\
{2}&{1.26}&{1.263}&{4.67$\times10^{-4}$}\\
{3}&{1.234}&{1.118}&{4.93$\times10^{-4}$}\\
{4}&{1.236}&{1.199}&{4.49$\times10^{-4}$}\\
\Xhline{3\arrayrulewidth}

\end{tabular}
\caption{Estimation of $\nu$ and $\lambda$ using method I}
\end{table}
\endgroup

\begin{figure}
\includegraphics[width=8cm]{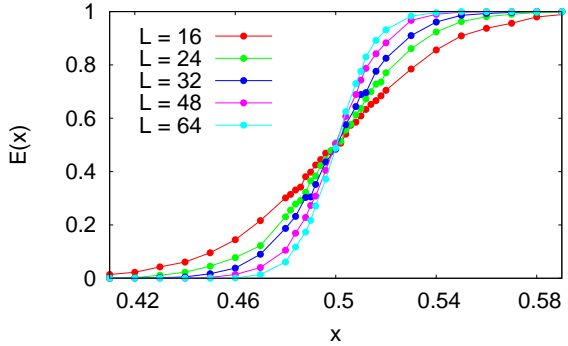}
 \caption{Unscaled data for exit probability for five different system sizes keeping $z = 0.5$}
\label{}
\end{figure}
\begin{figure}
\includegraphics[width=8cm]{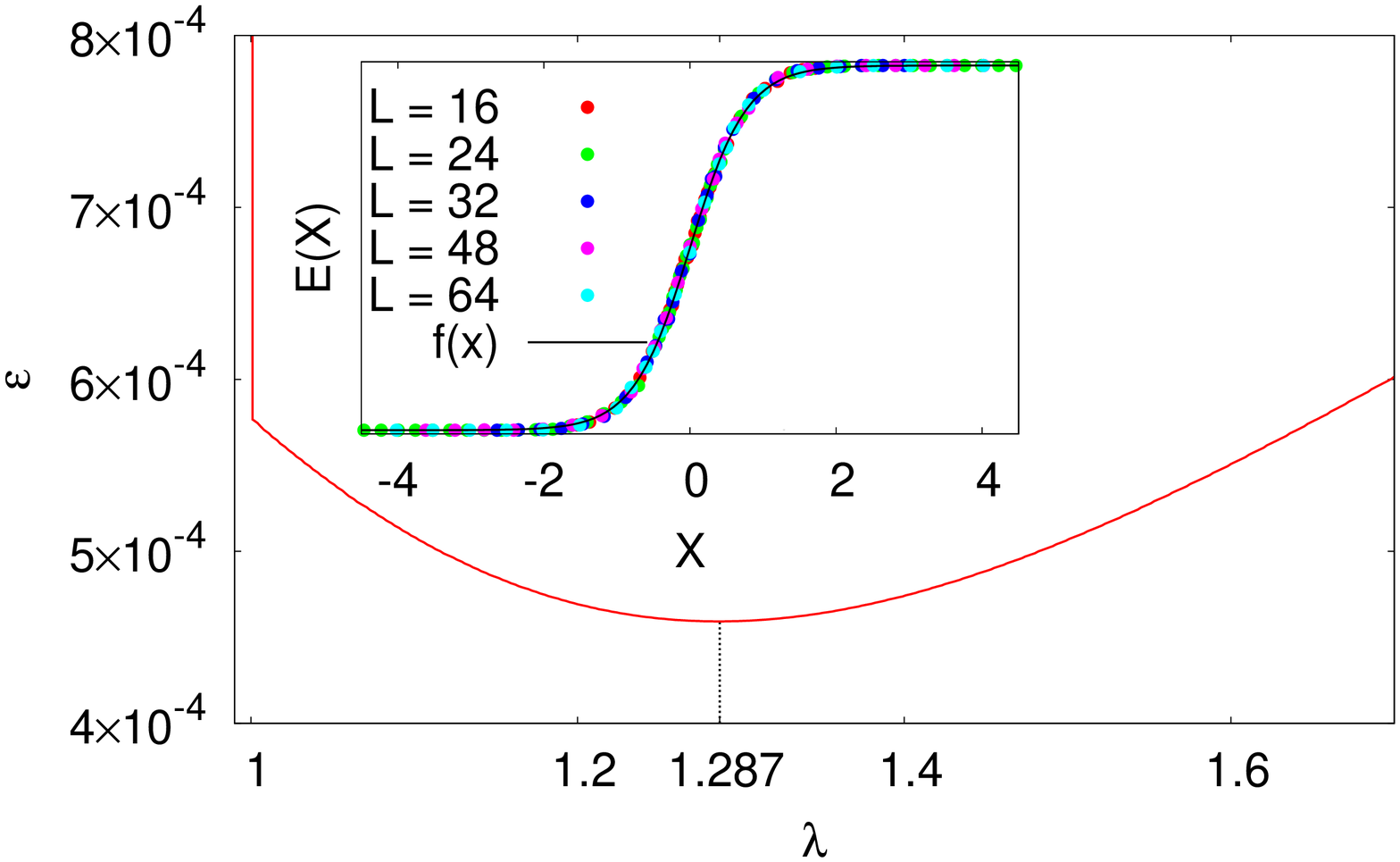}
 \caption{Variation of least square error (calculated by method I) with $\lambda$, for $\kappa = 0.5$ and $\nu = 1.29$ are shown. The minima of the curve is at $\lambda = 1.287$. Inset shows data collapse for exit probability for $\kappa = 0.5$ following method I. The collapse was done using $\nu = 1.29$ The collapsed plot was fitted according to $f(X) = [\tanh(\lambda X)+1]/2$, where $X = \big(\frac{x-x_c}{x_c}\big)L^{\frac{1}{\nu}}$, with $\lambda = 1.287$.}
\label{m1}
\end{figure}
\begin{figure}
\includegraphics[width=8cm]{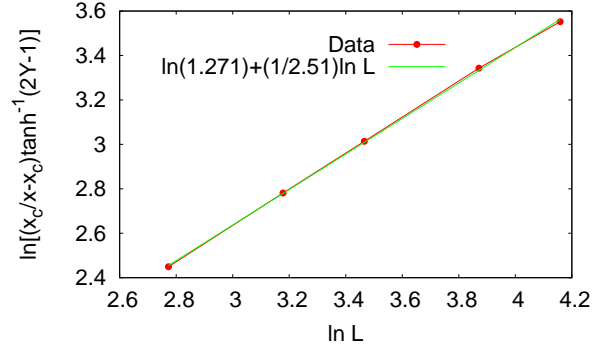}
 \caption{Plot of eq. (\ref{tanhscaled}) for $\kappa = 2$. The fitting straight line gives the value $\nu = 1.251$ and $\lambda = 1.271$}
\label{m2}
\end{figure}

\begingroup
\begin{table}

\begin{tabular}{|c|c|c|}
\Xhline{3\arrayrulewidth}
{$\kappa$}&{$\nu$}&{$\lambda$}\\
\Xhline{3\arrayrulewidth}
{0}&{1.269$\pm$0.038}&{1.124$\pm$0.093}\\
{0.1}&{1.235$\pm$0.049}&{1.187$\pm$0.13}\\
{0.25}&{1.294$\pm$0.048}&{1.355$\pm$0.136}\\
{0.5}&{1.236$\pm$0.022}&{1.162$\pm$0.059}\\
{0.75}&{1.231$\pm$0.026}&{1.189$\pm$0.069}\\
{0.8}&{1.209$\pm$0.048}&{1.121$\pm$0.123}\\
{0.9}&{1.237$\pm$0.026}&{1.195$\pm$0.069}\\
{1}&{1.289$\pm$0.011}&{1.339$\pm$0.031}\\
{1.1}&{1.196$\pm$0.032}&{1.115$\pm$0.086}\\
{1.5}&{1.329$\pm$0.014}&{1.516$\pm$0.044}\\
{2}&{1.251$\pm$0.014}&{1.271$\pm$0.039}\\
{3}&{1.236$\pm$0.008}&{1.227$\pm$0.023}\\
{4}&{1.188$\pm$0.099}&{1.213$\pm$0.037}\\
\Xhline{3\arrayrulewidth}
\end{tabular}
\caption{Estimation of $\nu$ and $\lambda$ using method II}
\end{table}
\endgroup

Finite size scaling analysis of the exit probability $E(x)$ was done according to 
\begin{equation}
E(x,L)=f\Bigg({\frac{(x-x_c)}{x_c}}L^{1/\nu}\Bigg),
\label{exiteq}
\end{equation}
where $f(y)\rightarrow0$ for $y<<0$ and $f(y) = 1$ for $y>>0$ (i.e. a step function like behaviour). The scaling form (\ref{exiteq}) which involves a dimensionless scaling variable $\big(\frac{x-x_c}{x_c}\big)L^{1/\nu}$ has been used earlier \cite{EPscaling,parnaEP,parna-arxiv,pratik} and very impressive data collapses have been obtained using it. This scaling argument indicates that $L^{-1/\nu}$ is basically the factor by which the width of the region, where $E(x)$ is not equal to 0 or 1, decreases. The scaling function $f$ fits with the form 
\begin{equation} 
f(X)=[\tanh(\lambda X)+1]/2,
\label{tanh}
\end{equation}
where $X = \big(\frac{x-x_c}{x_c}\big)L^{\frac{1}{\nu}}$.

Our intention was to calculate the exponents $\nu$ and $\lambda$ for the chosen values of $\kappa$. One could have simply done it by finding out the best collapsed curve using (\ref{exiteq}) and fit the scaled curve using (\ref{tanh}). But that would require us to depend largely on eye estimations to get the collapsed data. We have therefore used two different approaches which gives us a more unbiased estimate.

{\it{Method I:}} The data obtained for various system sizes were scaled using a rough value of $\nu$ according to Eq. (\ref{exiteq}). We next calculated $X = \big(\frac{x-x_c}{x_c}\big)L^{\frac{1}{\nu}}$ for chosen values of $L$. The range of $\nu$ was chosen as the approximate range within which the collapsed fit seemed fine visually. Now the $E(X)$ vs $X$ data is supposed to fit according to Eq. (\ref{tanh}). To perform the fitting we choose a suitable range of $\lambda$ and calculate $f(X)$ within this range using values of $\nu$ within this range. For every pair of $\nu$ and $\lambda$ we calculate the least square error defined as
\begin{equation}
\epsilon = \frac{1}{n}\sqrt{\sum_{n}(f(X) - E(X))^2}
\end{equation}
and for a given $\kappa$, the pair of $\nu$ and $\lambda$ which gave the minimum $\epsilon$ was chosen as the optimum ones.
In Table II, the values of $\nu$ and $\lambda$, and the corresponding least square error $\epsilon$ for every $\kappa$,
is recorded. In Fig. \ref{m1} method I has been demonstrated graphically for $\kappa=0.5$. The inset of Fig. \ref{m1} shows collapsed fit
for $E(x)$ for $\kappa=0.5$ using the exponents $\nu$ and $\lambda$ obtained by method I.

{\it{Method II:}} Combining Eq. (\ref{exiteq}) and (\ref{tanh}) we can write,
\begin{equation}
Y=E(X)=\frac{1}{2}[\tanh(\lambda X) + 1],
\end{equation}
And hence,
\begin{equation}
\ln\Big[\Big(\frac{x_c}{x-x_c}\Big) \tanh^{-1}(2Y-1)\Big] = \ln\lambda + \frac{1}{\nu} \ln L.
\label{tanhscaled}
\end{equation}

So, a plot of the left hand side of the Eq. (\ref{tanhscaled}) versus $\ln L$ would give a straight line. The straight line for $\kappa=2$ are shown as an example in Fig. \ref{m2}. From the slope and the intercept of the straight line we obtained the values of $\nu$ and $\lambda$.  

\begin{figure}
\centering
\includegraphics[width=9cm]{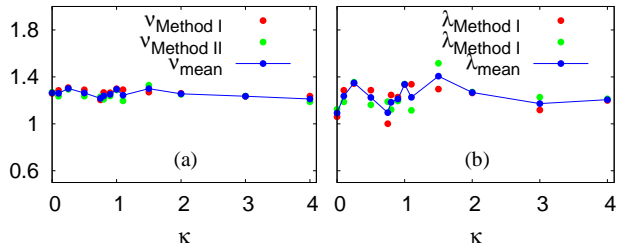}
 \caption{Variation of (a) $\nu$ and (b) $\lambda$ (b) with $\kappa$ obtained by Method I and Method II are shown. The solid line represents the mean value of the quantities obtained by the two methods at a particular $\kappa$.}
\label{nulamb}
\end{figure}

The exponents $\nu$ and $\lambda$ are apparently independent of $\kappa$, as obtained by both the methods (Fig. \ref{nulamb}); there is no systematic variation and the data indicate fluctuations about a mean value. Therefore, the mean values of $\nu$ and $\lambda$ for all $\kappa$
are taken and found to be $\bar{\nu} = 1.256$ (standard deviation $3.54\times 10^{-2}$) and $\bar{\lambda} = 1.231$ (standard deviation 0.11).

\section{Summary and conclusions}

In this paper, we presented numerical results for several features of the Ising model in two dimensions with Moore neighbourhood following a zero temperature quench. The interaction with the second nearest neighbours, located diagonally across, occurs with a strength $\kappa$ with respect to the nearest neighbour interaction.

Qualitatively, most of the results are similar to the nearest neighbour model barring a few exceptions. Frozen states occur for all $\kappa \geq 0$; however, the nature of the frozen states differs as one increases $\kappa$. For $\kappa = 0$, frozen states are comprised of striped patterns only, for $0 < \kappa \leq 2$, one encounters frozen states with either striped states or up and down domains separated by staircase like interfaces. For $\kappa>1$, another type of frozen state may also occur, namely, frozen states with local anti-ferromagnetic domains separating two domains with up or down spins. Such states have been observed only for $\kappa>2$, where they are more probable.

Our model has apparent similarity with the two-dimensional Ising model with $n$th neighbour interaction considered in \cite{prl-redner}; in both the models staircase like patterns occur as frozen states. However there are major differences, our model is clearly an extension of the nearest neighbour model to longer ranges in the minimum possible manner and is different from the case $n = 2$ (and obviously for all other $n$ values) in \cite{prl-redner}. So while staircase like patterns (subtly different in nature) are stable in both cases, we also obtain stable configurations with local anti-ferromagnetic motifs which can also be regarded as two adjacent staircase like interfaces (Fig. \ref{fstates810}(a)). This is actually due to the asymmetry in the interactions, a feature not considered earlier.

It is also found that for finite sizes, one may get iso-energetic active states for $\kappa>2$ as in three dimensions. These, however do not survive in the thermodynamic limit.

The freezing probability was studied as a function of $\kappa$ for different system sizes. As in \cite{spirin65,spirin63}, we find that the freezing probability increases with system size; however we do not find the tendency to decrease for larger sizes (we have checked up to $L = 96$) even for $\kappa = 0$. We also note that as a function of $\kappa$, the freezing probability shows an abrupt increase from the value at $\kappa = 0$, remains constant till $\kappa = 2$ where it decreases again to a constant for larger values of $\kappa$. However, the probabilities are not drastically different, they differ by $\sim 0.02$. We already note from Table \ref{table1} that results for all $\kappa>2$ should be identical, which is confirmed by this behaviour and of other quantities like fraction of spin flips, fraction of anti-ferromagnetic motifs etc.

In addition, we have attempted a study on the variation of persistence probability with time. The results do not differ drastically for homogeneous initial conditions as far as the persistence exponent is concerned, but are clearly different for the non-homogeneous case. For $x=0.5$ (i.e. homogeneous initial condition), although $\theta_0$ and $\theta_{\kappa}$ are different even when the error bars are considered, they are within ten percent of each other. The most accurate estimate for $\theta_0$ is believed to be 0.199 $\pm$ 0.002 \cite{blanchard} which also differs by about ten percent from our results. Hence it could be interpreted that the exponents $\theta_0$ and $\theta_k$ are not different i.e. the persistence exponent for $x=0.5$ is $\kappa$ independent.

For $x\ne 0.5$ and $\kappa>0$, $P_{min}$ shows an algebraic decay followed by a stretched exponential cutoff but with different values of the exponents compared to $\kappa=0$. $P_{maj}$ and $P_{total}$ for $\kappa>0$ saturate at higher values than $\kappa=0$. Hence, the addition of the second nearest neighbours has resulted in a different universality class as far as the persistence behaviour for non-homogeneous system initial condition is concerned. More drastic variation of the persistence exponent was found in \cite{chandra-sdg}.

The results obtained for the exit probability $E(x)$ are also not different from that of the nearest neighbour case.
Both the exponents $\bar{\nu}\approx 1.26$ and $\bar{\lambda}\approx 1.23$ associated with the scaling of $E(x)$ show universal
behaviour.
It may be mentioned here that in a previous work by the present authors \cite{pratik} the estimate of $\nu$ was obtained as $\sim 1.47$;
however, it has been checked that the overestimation occurred as lesser number of points were considered close
to $x = 0.5$. In this particular work,
two different methods were used to obtain $\nu$ and $\lambda$ and we believe that the present result is more accurate than that in \cite{pratik}. Another recent work \cite{parna-arxiv} confirms that $\nu$ is indeed close to 1.3. 

Considering all the results, one may conclude that the Moore neighbourhood and the asymmetry in the interactions are
effective as far as the freezing behaviour and persistence phenomena for non-homogeneous initial condition are considered, while the behaviour of exit probability remains the same. Considering the Moore neighbourhood is the minimum possible way the range of a two dimensional nearest neighbour model can be extended. We therefore get interesting effects of the minimal extension, however, the effects are most prominent when $\kappa>2$. In fact, a sharp transition is obtained at $\kappa = 2$ when one considers iso-energetic active states (only for finite sizes), freezing probability and fraction of anti-ferromagnetic motifs. However, no such change, either smooth or sharp, is observed for other quantities like the exponents for exit probability and persistence probability.

Of course, switching
off the nearest neighbour interaction altogether (i.e. making $\kappa \to \infty$) will split the lattice into two independent
two dimensional nearest neighbour Ising models. However considering this limit is beyond the scope of the present work as $J=1$ has been used throughout.

Acknowledgement: P. Mullick thanks DST-INSPIRE (Sanction No. 2015/IF0673) for financial support. P. Sen thanks CSIR (Government of India) for financial grant. Discussions with S. Dasgupta and A. Banerjee is also acknowledged.


\begin{thebibliography}{99}

\bibitem{bray}
A. J. Bray, Adv. Phys. \textbf{51} (2), 481 (2002).

\bibitem{krapbook}
P. L. Krapivsky, S. Redner, and Eli Ben-Naim, \textit{A Kinetic View
of Statistical Physics} (Cambridge University Press, Cambridge,
2010)

\bibitem{corberi}
On lattices however it has been argued that the mechanism might be different which affects the results for the zero temperature coarsening in three dimensions. See F. Corberi, E. Lippiello, M. Zannetti, Phys. Rev. E \textbf{78}, 011109 (2008).

\bibitem{halperin}
P. C. Hohenberg and B. I. Halperin, Rev. Mod. Phys. \textbf{49}, 435 (1977).

\bibitem{lipowski}
A. Lipowski, Physica A \textbf{268}, 6 (1999).

\bibitem{spirin65}
V. Spirin, P. L. Krapivsky and S. Redner, Phys. Rev. E \textbf{65}, 016119 (2001).

\bibitem{spirin63}
V. Spirin, P. L. Krapivsky and S. Redner, Phys. Rev. E \textbf{63}, 036118 (2001).

\bibitem{barros}
K. Barros, P. L. Krapivsky and S. Redner, Phys. Rev. E \textbf{80}, 040101(R) (2009).

\bibitem{prl-redner}
J. Olejarz, P. L. Krapivsky and S. Redner, Phys. Rev. Lett. \textbf{109}, 195702 (2012).

\bibitem{derrida1}
B. Derrida, A. J. Bray, and C. Godreche, J. Phys. A \textbf{27}, L357 (1994).

\bibitem{satya3}
A. J. Bray, S. N. Majumdar, and G. Schehr, Adv. Phys. \textbf{62}, 225 (2013).

\bibitem{stauffer1}
D. Stauffer, J. Phys. A: Math. Gen. \textbf{27}, 5029 (1994).

\bibitem{derrida2}
B. Derrida, J. Phys. A \textbf{28}, 1481 (1995).

\bibitem{bennaim}
E. Ben-Naim, L. Frachebourg, and P. L. Krapivsky, Phys. Rev. E \textbf{53}, 3078 (1996).

\bibitem{jain}
S. Jain, Phys. Rev. E \textbf{59}, R2493 (1999).

\bibitem{blanchard}
T. Blanchard, L. F. Cugliandolo, and M. Picco, J. Stat. Mech. (2014) P12021.

\bibitem{subir1}
S. Chakraborty and S. K. Das, Eur. Phys. J. B \textbf{88}, 160 (2015).

\bibitem{subir2}
S. Chakraborty and S. K. Das, Phys. Rev. E, \textbf{93}, 032139 (2016).

\bibitem{pratik}
P. Mullick and P. Sen, Phys. Rev. E \textbf{93}, 052113 (2016).

\bibitem{parna-arxiv}
P. Roy and P. Sen, Phys. Rev. E \textbf{95}, 020101(R) (2017).

\bibitem{redkrap}
S. Redner and P. L. Krapivsky, J.Phys. A \textbf{31}, 9229 (1998).

\bibitem{das-barma}
D. Das and M. Barma, Physica A \textbf{270}, 245 (1999); Phys. Rev. E \textbf{60}, 2577 (1999).

\bibitem{ps-sdg}
P. Sen and S. Dasgupta, J. Phys. A \textbf{37}, 11949 (2004).

\bibitem{chandra}
S. Biswas, A. K. Chandra and P. Sen,  Phys. Rev. E \textbf{78}, 041119 (2008).

\bibitem{chandra-sdg}
A. K. Chandra and S. Dasgupta, Phys. Rev. E \textbf{77}, 031111 (2008).

\bibitem{castellano}
C. Castellano and R. Pastor-Satorras, Phys. Rev. E \textbf{83}, 016113 (2011).

\bibitem{parnajstat}
P. Roy and P. Sen, J. Stat. Phys. \textbf{159}, 893904 (2105). 

\bibitem{parnasoham}
P. Roy, S. Biswas, P. Sen, Phys. Rev. E \textbf{89}, 030103(R) (2014).

\bibitem{cwli}
C. W. Li et al, Phys. Rev. B \textbf{90}, 214303 (2014).


\bibitem{newman}
M. E. J. Newman and G. T. Barkema, \textit{Monte Carlo Methods in Statistical Physics}, Oxford University Press (1999).

\bibitem{EPscaling}
S. Biswas, S. Sinha and P. Sen, Phys. Rev. E \textbf{88}, 022152 (2013).

\bibitem{parnaEP}
P. Roy, S. Biswas and P. Sen, J. Phys. A: Math. Theor. \textbf{47}, 495001 (2014).
















\end{thebibliography}
\end{document}